\documentclass[conference]{IEEEtran}

\IEEEoverridecommandlockouts

\usepackage[english]{babel}
\ifCLASSINFOpdf
  \usepackage[pdftex]{graphicx}					
  \DeclareGraphicsExtensions{.pdf,.jpeg,.png,.eps}
\else
  \usepackage[dvips]{graphicx}
  \DeclareGraphicsExtensions{.eps}
\fi

\usepackage[caption=false]{subfig}	
\graphicspath{{Figures/}}
\usepackage{ifpdf}
\ifpdf
  \usepackage{epstopdf}
\fi

\usepackage{amsmath, amsthm, amssymb}
\usepackage{bbm, amsfonts}

\usepackage{algorithm,algorithmic}
\usepackage{textcomp}

\usepackage{cite}
\usepackage{cleveref}
\usepackage{xcolor,tikz}

\crefname{figure}{fig.}{figs.}

\crefname{app}{Appendix}{Appendices}
\crefname{cor}{Corollary}{Corollary}
\crefname{prop}{Proposition}{Proposition}
\crefname{lemma}{Lemma}{Lemma}
\crefname{defn}{Definition}{Definition}
\crefname{conj}{Conjecture}{Conjecture}
\crefname{exam}{Example}{Example}
\crefname{supp}{Supplemental Section}{Supplemental Section}

\newcommand{\bs}{\boldsymbol}
\newcommand{\bb}{\mathbb}
\newcommand{\mcal}{\mathcal}

\newcommand{\lb}{\left(}
\newcommand{\rb}{\right)}
\newcommand{\ls}{\left[}
\newcommand{\rs}{\right]}
\newcommand{\lc}{\left\{}
\newcommand{\rc}{\right\}}

\newcommand{\lv}{\left\vert}
\newcommand{\rv}{\right\vert}

\newcommand{\LRV}[1]{{\left\vert\kern-0.25ex\left\vert\kern-0.25ex\left\vert #1 \right\vert\kern-0.25ex\right\vert\kern-0.25ex\right\vert}}

\newcommand{\expect}[2]{\bb{E}_{#1}\lc#2\rc}

\newcommand{\nth}{^\mathsf{th}}

\newcommand{\bbP}{\bb{P}}

\newcommand{\calA}{\mcal{A}}

\newcommand{\calE}{\mcal{E}}

\newcommand{\calP}{\mcal{P}}

\newcommand{\vecs}{\bs{s}}

\newcommand{\vecy}{\bs{y}}

\makeatletter
\def\ps@IEEEtitlepagestyle{%
  \def\@oddfoot{\mycopyrightnotice}%
  \def\@evenfoot{}%
}
\def\mycopyrightnotice{%
  {\footnotesize 978-1-7281-5478-7/20/\$31.00~\copyright2020 IEEE \hfill}
  \gdef\mycopyrightnotice{}
}

\begin{document}
\title{Anomaly Detection Under Controlled Sensing Using Actor-Critic Reinforcement Learning\\
\thanks{This work was supported in part by the  National Science Foundation under { grants ENG 60064237 and CCF 1618615}.
}}
\author{\IEEEauthorblockN{Geethu Joseph, M. Cenk Gursoy, and Pramod K. Varshney}
\IEEEauthorblockA{\textit{Department of of Electrical Engineering and Computer Science } \\
\textit{Syracuse University}\\
 New York, USA}
Emails:\{gjoseph,mcgursoy,varshney\}@syr.edu.}

\maketitle

\begin{abstract}
We consider the problem of detecting anomalies among a given set of processes using their noisy binary sensor measurements. The noiseless  sensor measurement corresponding to a normal process is 0, and the measurement is 1 if the process is anomalous. The decision-making algorithm is assumed to have no knowledge of the number of anomalous processes. The algorithm is allowed to choose a subset of the sensors at each time instant until the confidence level on the decision exceeds the desired value. Our objective is to design a sequential sensor selection policy that dynamically determines which processes to observe at each time and when to terminate the detection algorithm. The selection policy is designed such that the anomalous processes are detected with the desired confidence level while incurring  minimum cost which comprises the delay in detection and the cost of sensing. We cast this problem as a sequential hypothesis testing problem within the framework of Markov decision processes, and solve it using the actor-critic deep reinforcement learning algorithm. This deep neural network-based algorithm offers a low complexity solution with good detection accuracy. We also study the effect of statistical dependence between the processes on the algorithm performance. Through numerical experiments, we show that our algorithm is able to adapt to any unknown statistical dependence pattern of the processes.
\end{abstract}
\section{Introduction}
The anomaly detection problem in this paper refers to the estimation of states of $N$ given (not necessarily independent) processes. Each process can be in two states, either normal or anomalous. Also, each process is monitored using a noisy binary sensor. The sensor may observe the state of the process incorrectly, i.e., its observed state is flipped from the actual state, with a certain probability. In this context, the notion of controlled (active) sensing is the ability of the decision-making agent to adaptively control the observations of the system by switching between various sensor subsets~\cite{krishnamurthy2016partially}, i.e., the decision-maker (detection algorithm) chooses a potentially different  subset of sensors at each time instant to determine the states of the processes. One motivating application for our problem is remote system health monitoring using a wireless sensor network (see, for example, \cite{chung2006remote,bujnowski2013enhanced}). Each sensor in the network observes a different (but not necessarily independent) functionality of the system and sends the collected data to the monitoring center over a wireless channel. Due to the uncertainty introduced by  the channel conditions, the received data can be corrupted with some probability.  In this setting, taking additional measurements yields more accurate estimates but this incurs a higher energy consumption, and this, in turn, reduces the life span of the sensor network. On the other hand, making fewer measurements leads to a larger delay in identifying a potential system malfunction. Therefore, it is important to design a sequential sensor selection policy that can reliably infer the state of the processes as quickly as possible by using as few sensor measurements as possible.

\subsection{Related literature}
The anomaly detection problem considered in this paper is a special case of active hypothesis testing that dates back to 1959~\cite{chernoff1959sequential}. The goal of active hypothesis testing is to deduce whether one of several hypotheses is true by gathering relevant  data. The decision-making algorithm performs experiments until sufficiently strong evidence is gathered. In \cite{chernoff1959sequential}, the author  proposed a randomized strategy and established its asymptotic optimality. However, the solution involved solving an optimization problem at each time and thus, it is computationally expensive. This seminal work in~\cite{chernoff1959sequential} was followed by several other studies that investigated active hypothesis testing  under different settings~\cite{bessler1960theory,nitinawarat2013controlled,naghshvar2013active,huang2018active}. These papers characterized the theoretical aspects of the problem and presented a few model-based algorithms to solve the problem. Recently, some works have explored deep neural network based-learning algorithms for active hypothesis testing~\cite{kartik2018policy,zhong2019deep}. These deep learning-based approaches have  provided low complexity algorithms that are practically useful. However, these studies do not incorporate the cost of sensing in the detection problem and assume that the  decision-maker chooses the same fixed number of sensors at every time instant. Our problem setting is different from these models. Specifically, we consider the case where the decision-maker can choose any number of sensors at each time instant, and this choice is determined by the cost associated with the sensor measurements. Additionally, we take into account the potential dependence among different processes.

\subsection{Our contributions}
We formulate, in \Cref{sec:mdp}, the anomaly detection problem as a Markov decision process (MDP) problem. In \Cref{sec:actorcritic}, we present the actor-critic framework that learns an optimal policy that dynamically selects the sensors at each time instant by minimizing the cost of sensing subject to the condition that the confidence level on the estimated states of the processes exceeds a specific value. We note that if the states of any two processes are dependent, the sensor measurement corresponding to one process provides information about the other process impacting the overall system operation and performance. In \Cref{sec:simulations}, through numerical simulation, we observe that when the dependence between the states of the processes is high, the delay in state estimation is small. This result implies that the algorithm is able to learn any underlying statistical dependence among the processes and reduce the number of sensor measurements by taking advantage of this dependence.

In summary, we present a low-complexity algorithm based on the actor-critic method for the anomaly detection problem and study the effect of statistical dependence between the processes and the cost of sensing on the algorithm performance.
\section{Anomaly Detection Under Controlled Sensing}\label{sec:system}
We consider $N$ processes where each of the processes is in one of the two states: normal (denoted by 0) or anomalous (denoted by 1). The states of the processes are denoted by $\vecs\in\{0,1\}^N$ where { the $i\nth$ entry $\vecs_i$ is the state of the $i\nth$ process.}

Each process is monitored by a sensor, and the sensor measurement corresponding to the $i\nth$ process at time instant $k$ is denoted by $\vecy_i(k)\in\{0,1\}$. The uncertainty or potential error in the noisy measurement is modeled using a binary symmetric channel with a cross-over probability $p$:
\begin{equation}\label{eq:mesurement}
\vecy_{i}(k) = \begin{cases}
{\vecs_{i}} & {\text{ with probability } 1-p,}\\
1-{\vecs_{i}}& {\text{ with probability } p}.
\end{cases}
\end{equation}
For any integer $K>0$, given the  state vector {$\vecs$}, the measurements $\lc\vecy_i(k),i\in[N],k\in[K]\rc$ are jointly (conditionally) independent. Here, the notation $[\cdot]$ is defined as
$[a]\triangleq\{1,2,\ldots,a\},$
for any positive integer $a$. The cost associated with each sensor measurement is denoted by $\lambda>0$. The goal of this work is to find the optimal (in terms of cost of sensing) sensor selection policy  so that the time required to estimate the states of the processes is minimum while yielding detection with desired confidence. We note that the number of anomalous processes is also unknown to the decision-maker. To find the optimal policy, we cast our anomaly detection problem into an active hypothesis testing framework in the next section.

\section{Active hypothesis testing}\label{sec:mdp}
For $N$ processes, there are $2^N$ possible values for the state vector $\vecs$. Similarly, at every time instant, the algorithm can pick any number of sensors and thus, there are $2^N-1$ possible actions. We omit the action where the decision-maker does not choose any sensor as we assume that the algorithm collects data until it makes a decision.  Hence, the active  hypothesis testing problem that is equivalent to the anomaly detection problem has $2^N$ hypotheses and $2^N-1$ possible actions. Next, we formulate an infinite-horizon, average-reward MDP problem that solves this active hypothesis testing problem.

The state of the MDP is the posterior belief $\pi\in[0,1]^{2^N}$ on the set of all possible hypotheses. Let the sequence of actions selected by the decision-maker be $\lc\calA_k\subseteq[N],k=1,2,\ldots\rc$. Also, let $H\in[2^N]$ be the true hypothesis, and $q_i$ be the prior probability that hypothesis $i$ is true. Using the available information,  the decision-maker computes a posterior belief vector $\pi(k)\in[0,1]^{2^N}$ at time $k$ whose $i\nth$ entry is given by
{\begin{align}
\pi_i(k) &= \bbP\lb H=i\middle| \calA_j,j=1,2,\ldots,k\rb\\
&= \frac{\bbP\lb H=i\rb\bbP\lb \vecy_{\calA_j}(j) ,j=1,2,\ldots,k\middle| H=i \rb}{\sum_{i'} \bbP\lb H=i'\rb\bbP\lb \vecy_{\calA_j}(j) ,j=1,2,\ldots,k\middle| H=i' \rb}\\
&= \frac{q_i \prod_{j=1}^k \prod_{a\in\calA_j} \ls (1-p) \mathbbm{1}_{ \calE_{a,j,i}}+ p\mathbbm{1}_{\calE_{a,j,i}^c} \rs}{\sum_{i'} q_{i'} \prod_{j=1}^k \prod_{a\in\calA_j}  \ls (1-p) \mathbbm{1}_{ \calE_{a,j,i'}}+ p\mathbbm{1}_{\calE_{a,j,i'}^c}\rs}\label{eq:posterior}\\
&= \frac{\pi_i(k-1) \prod_{a\in\calA_k}\ls (1-p) \mathbbm{1}_{\calE_{a,k,i}}+ p\mathbbm{1}_{\calE_{a,k,i}^c} \rs}{\sum_{i'} \pi_{i'}(k-1) \prod_{a\in\calA_k} \ls (1-p) \mathbbm{1}_{\calE_{a,k,i'}}+ p\mathbbm{1}_{\calE_{a,k,i'}^c}\rs},\label{eq:posterior_update}
\end{align}}
where $\mathbbm{1}$ is the indicator function and $\calE_{a,j,i}$ denotes the event that the sensor measurement and the corresponding state are the same:
\begin{equation}
\calE_{a,j,i}\triangleq \lc \vecy_a(j)=\vecs_a\middle| H=i\rc.
\end{equation}
The event $\calE_{a,j,i}^c$ denotes the complement of $\calE_{a,j,i}$. {Further, \eqref{eq:posterior} follows from the conditional independence of the measurements, given the state vector. To get  \eqref{eq:posterior}, we also use \eqref{eq:mesurement} which gives $\bbP\lc\calE_{a,j,i}\rc=1-p$ for all values of $a,i$ and $j$.}

The Bayesian log-likelihood ratio of hypothesis $i\in[2^N]$ at time $k$ is given by
\begin{equation}\label{eq:Cdefn}
C_i(\pi) = \log\frac{\pi_i}{1-\pi_i},
\end{equation}
where $\pi_i$ is the $i\nth$ entry of a posterior belief vector $\pi$.
The quantity $C_i(\pi)$ serves as a measure of confidence on hypothesis $i$ being true. Therefore, our goal is to find a sensor selection policy $\mu$ to increase the confidence level $C_H(\pi)$ on the true hypothesis $H$ as quickly as possible while keeping the sensing cost low. The policy $\mu:[0,1]^{2^N}\to\calP([N])\setminus\{\phi\}$ is a mapping from the posterior distribution $\pi$ to the action space whose elements are subsets of $[N]$. Here, $\calP([N])$ and $\phi$ denote the power set  of $[N]$ and the null set, respectively.

Inspired by the reward functions used in~\cite{kartik2018policy,zhong2019deep}, we define the objective function of MDP to be maximized as
\begin{align}\label{eq:accumulated_reward}
{R(K)} &=  \frac{1}{K}\expect{\mu}{  C_H(\pi(k))- C_H(\pi(0)) - \lambda\sum_{k=1}^K\lv\calA_k\rv}\\
&= \frac{1}{K}\lb  \expect{\mu}{\bar{C}(\pi(k))- \bar{C}(\pi(0))} - \lambda\sum_{k=1}^K\lv\calA_k\rv\rb,
\end{align}
where $\expect{\mu}{\cdot}$ is the expectation under policy $\mu$ and $K$ denotes the stopping time. We recall that $\lambda$ is the cost per sensor measurement and $\lc\calA_k\subseteq[N]\rc_{k=1}^K$ is the sequence of actions chosen by the decision-maker. We define the average Bayesian log likelihood ratio~\cite{kartik2018policy} $\bar{C}(\cdot)$ as
\begin{equation}\label{eq:confi}
\bar{C}(\pi) = \sum_{i=1}^{2^N} \pi_i C_i(\pi)=\sum_{i=1}^{2^N} \pi_i \log\frac{\pi_i}{1-\pi_i}.
\end{equation} Thus, the instantaneous reward of MDP is given by 
\begin{equation}\label{eq:imm_reward}
r(k) = \bar{C}(\pi(k)) - \bar{C}(\pi(k-1))- \lambda\lv\calA_k\rv.
\end{equation}
The objective in the MDP problem is to find the sequence of actions $\lc\calA_k\subseteq[N]\rc$ that maximizes the long-run average of the rewards: $\lim_{K\to\infty}\frac{1}{K}\sum_{k=1}^K \expect{g}{r(k)}$ which is the same as $\lim_{K\to\infty} {R(K)}$, as defined in \eqref{eq:accumulated_reward}. This MDP is solved using the actor-critic reinforcement learning approach as discussed in the next section.

\section{Actor-Critic Framework}\label{sec:actorcritic}
The actor-critic algorithm is designed for a discounted reward MDP formulation. So we first convert our average reward formulation to a discounted reward formulation with a discount factor $0<\gamma <1$ (which is close to 1) and the total discounted reward is defined as follows~\cite{sutton2018reinforcement}:
\begin{equation}
\lim_{K\to\infty}\frac{1}{K}\sum_{k=1}^K \gamma^{k-1} \expect{\mu}{r(k)}.
\end{equation}
The actor-critic architecture consists of two neural networks, namely, actor and critic. The actor learns the policy which chooses the action based on the posterior probabilities $\pi$. The critic estimates the value function which is an estimate of how good the policy learned by the actor  is and hence essentially provides an evaluation of that policy. The actor updates the policy based on the value function computed by the critic.

{The output layer of the actor network has $2^N-1$ nodes representing the set of all possible actions $\calP([N])\setminus\{\phi\}$. The value of each node is the probability of obtaining the maximum reward when the corresponding action is chosen. At every time instant $k$, the actor chooses action $\calA_k$ as $\calA_k=\mu(\pi(k-1))=\underset{a\in \calP([N])\setminus\{\phi\}}{\arg\max}\nu_{\theta}(a|\pi(k-1))$, where $\theta$ is the set of parameters of the actor network and $\nu_{\theta}(a|\pi(k-1))$ denotes the network output when $\pi(k-1)$ is fed as the input to the network.}  Thus, the decision-maker receives the corresponding observations denoted by $\vecy_{\calA_k,k}$. Then, based on the available information $\calA_k$ and $\vecy_{\calA_k,k}$, the posterior probability $\pi(k-1)$ is updated to  $\pi(k)$ using \eqref{eq:posterior_update}. We note that the knowledge of the crossover probability $p$ and the prior on the hypotheses are required for updating $\pi$. These parameters can easily be estimated from the training data.
Then, the instantaneous reward $r(k)$ is computed using  \eqref{eq:imm_reward} and it is fed to the critic along with the posterior probability pair, $\pi(k)$ and $\pi(k-1)$. The critique takes the form of temporal error $\delta$ as follows:
\begin{equation}\label{eq:temporalerror}
\delta(k) = r(k)+\gamma V(\pi(k))-V(\pi(k)),
\end{equation}
where the function $V$ is the current value function learned by the critic. This error $\delta(k)$ is used to evaluate the action $\calA_{k}$ selected by the actor for the posterior probability $\pi(k-1)$. The critic updates its neural network weights by minimizing the square of the temporal error $\delta^2(k)$ with respect to $V$. The actor also updates { its  parameter $\theta$} via the policy gradient using the temporal error computed by the critic~\cite[Chapter 13]{sutton2018reinforcement}:
{\begin{equation}\label{eq:policy_gradient}
\theta = \theta + \delta(k)\nabla_{\theta} \ls \log \nu_{\theta}(\calA_k|\pi(k-1))\rs.
\end{equation}
where $\nabla_{\theta}$ denotes the gradient with respect to $\theta$.}

\begin{algorithm}[t]
\caption{Actor-critic reinforcement learning for anomaly detection}
\label{alg:ActorCritic}
\begin{algorithmic}[1]
\REQUIRE

\begin{itemize}
\item []
\item Discount rate $\gamma\in(0,1)$
\item Maximum number of episodes $E_{\max}$
\item Maximum number of time slots $T_{\max}$
\item Upper threshold on confidence $\pi_{\mathrm{upper}}\in(0.5,1]$
\end{itemize}

\ENSURE
\begin{itemize}
\item []
\item Actor and critic neural networks with random weights
\item $\pi(0)$ with prior on the hypothesis (can be  learned from the training data)
\end{itemize}

\FOR {Episode index = 1 to $E_{\max}$}
\STATE Time index $k =0$
\WHILE {$\underset{i}{\max} \; \pi_i > \pi_{\mathrm{upper}}$ and $k<T_{\max}$}
\STATE Choose action $\calA_k=\mu(\pi(k))$
\STATE Generate sensor measurements $\vecy_{\calA_k,k}$
\STATE Compute $\pi(k+1)$ using \eqref{eq:posterior_update}
\STATE Compute instantaneous reward $r(k)$ using \eqref{eq:imm_reward}
\STATE Update the critic neural network by minimizing the temporal error $\delta(k)^2$ in \eqref{eq:temporalerror} with respect to $V$
\STATE {Update the actor neural network using \eqref{eq:policy_gradient}}
\STATE Increase time index $k=k+1$
\ENDWHILE
\STATE Declare the estimated hypothesis as $\underset{i}{\arg\max} \; \pi_i$
\ENDFOR
\end{algorithmic}
\end{algorithm}
The algorithm stops sensing and returns the hypothesis estimate when the confidence level exceeds the desired level. However, we note that the confidence $C_i(\pi)$ defined in {\eqref{eq:Cdefn}} is an increasing function of $\pi_i$. Hence, we define our desired level of confidence in terms of the desired level on the belief $\pi_i$. Therefore, the algorithm terminates when $\underset{i}{\max} \; \pi_i > \pi_{\mathrm{upper}}$, where $\pi_{\mathrm{upper}}$ is the desired level. The pseudo-code of our algorithm is given in \Cref{alg:ActorCritic}.
\section{Numerical Experiments}\label{sec:simulations}

\begin{figure*}
\begin{center}
\hspace{-0.8cm}
\subfloat[$\rho=0$]
{\includegraphics[width=6.8cm]{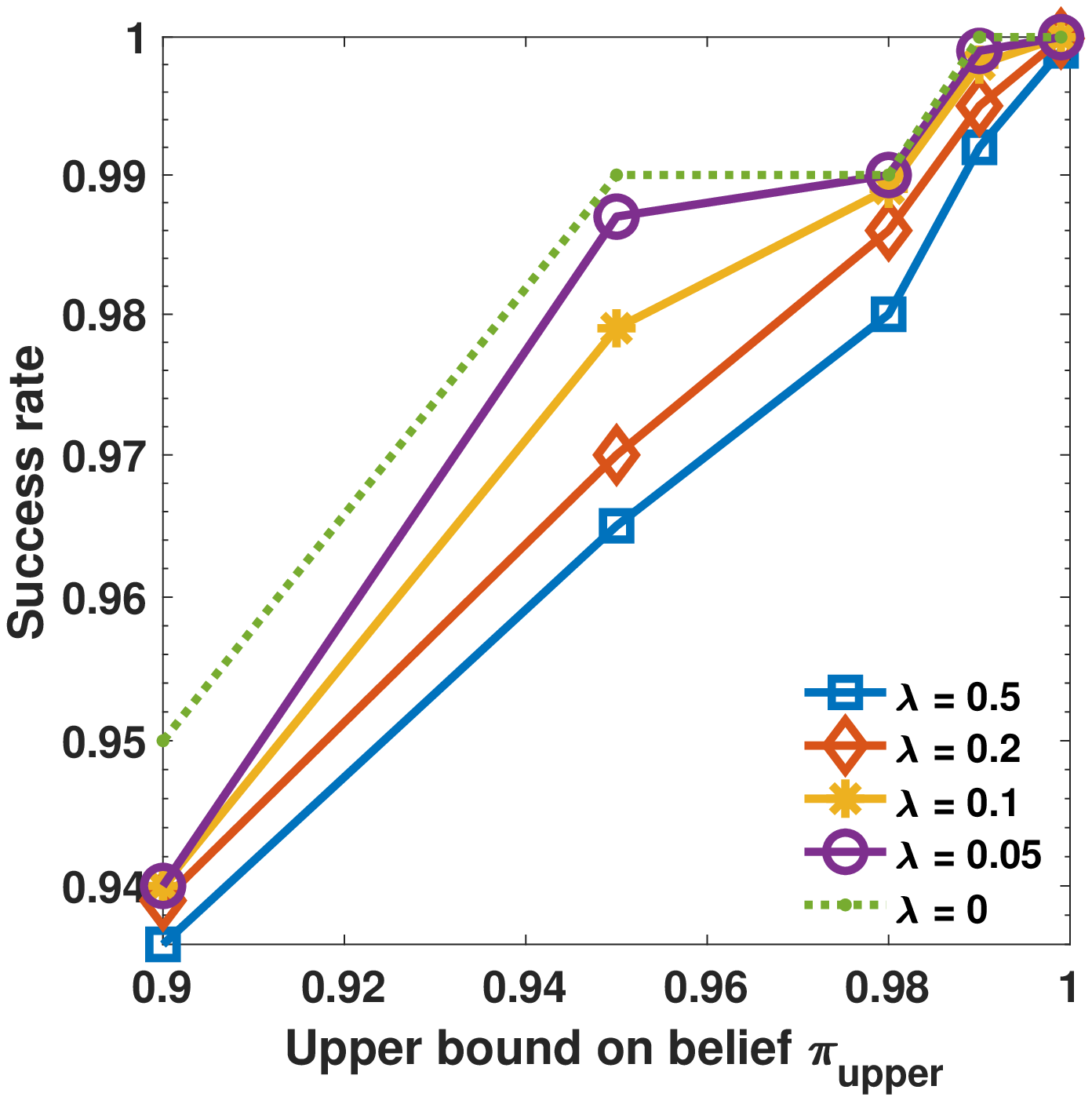}
\label{fig:success0}}
\hspace{-1.4cm}
\subfloat[$\rho=0.3$]
{\includegraphics[width=6.8cm]{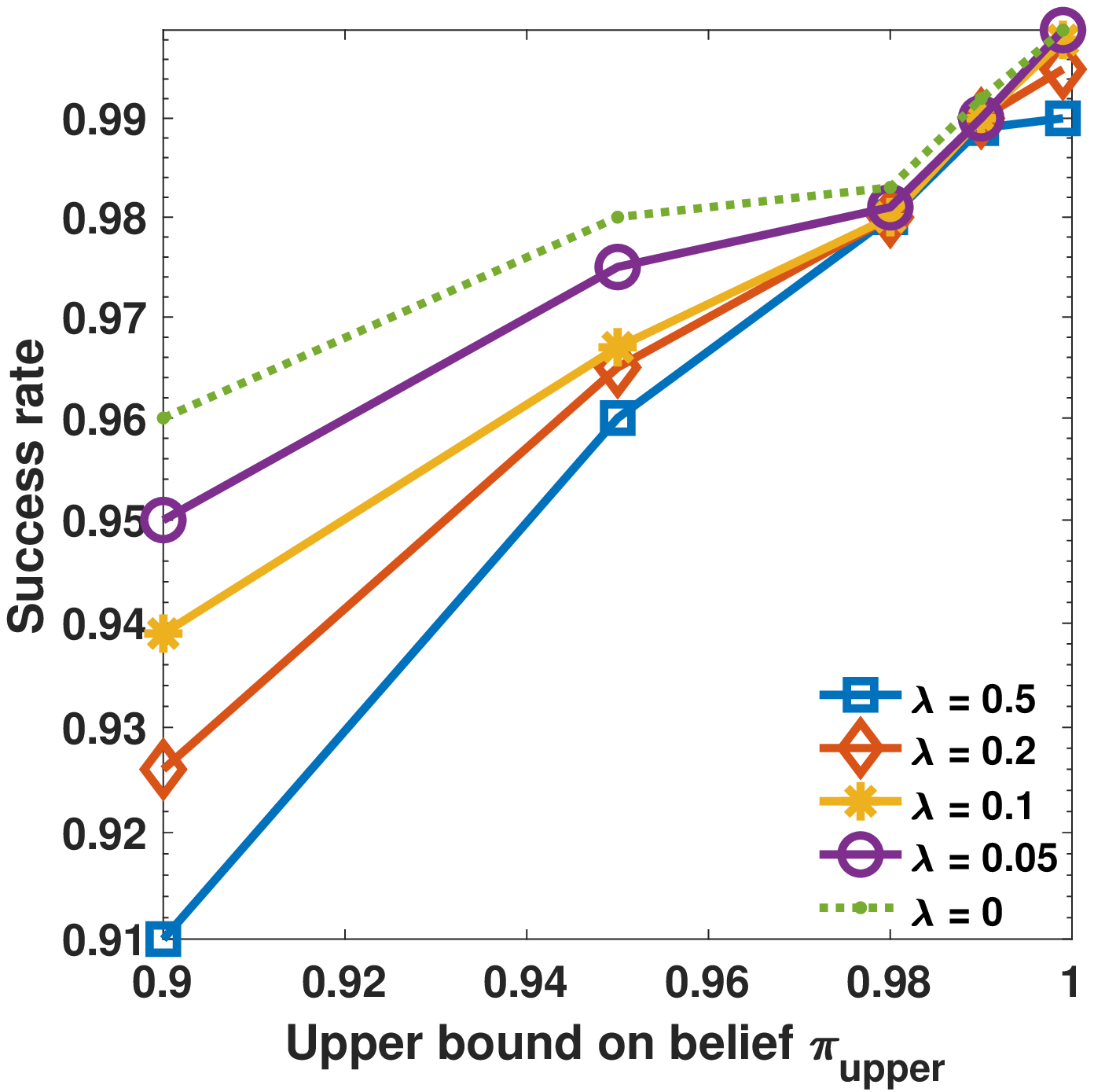}
\label{fig:success3}}
\hspace{-1.4cm}
\subfloat[$\rho=1$]
{\includegraphics[width=6.8cm]{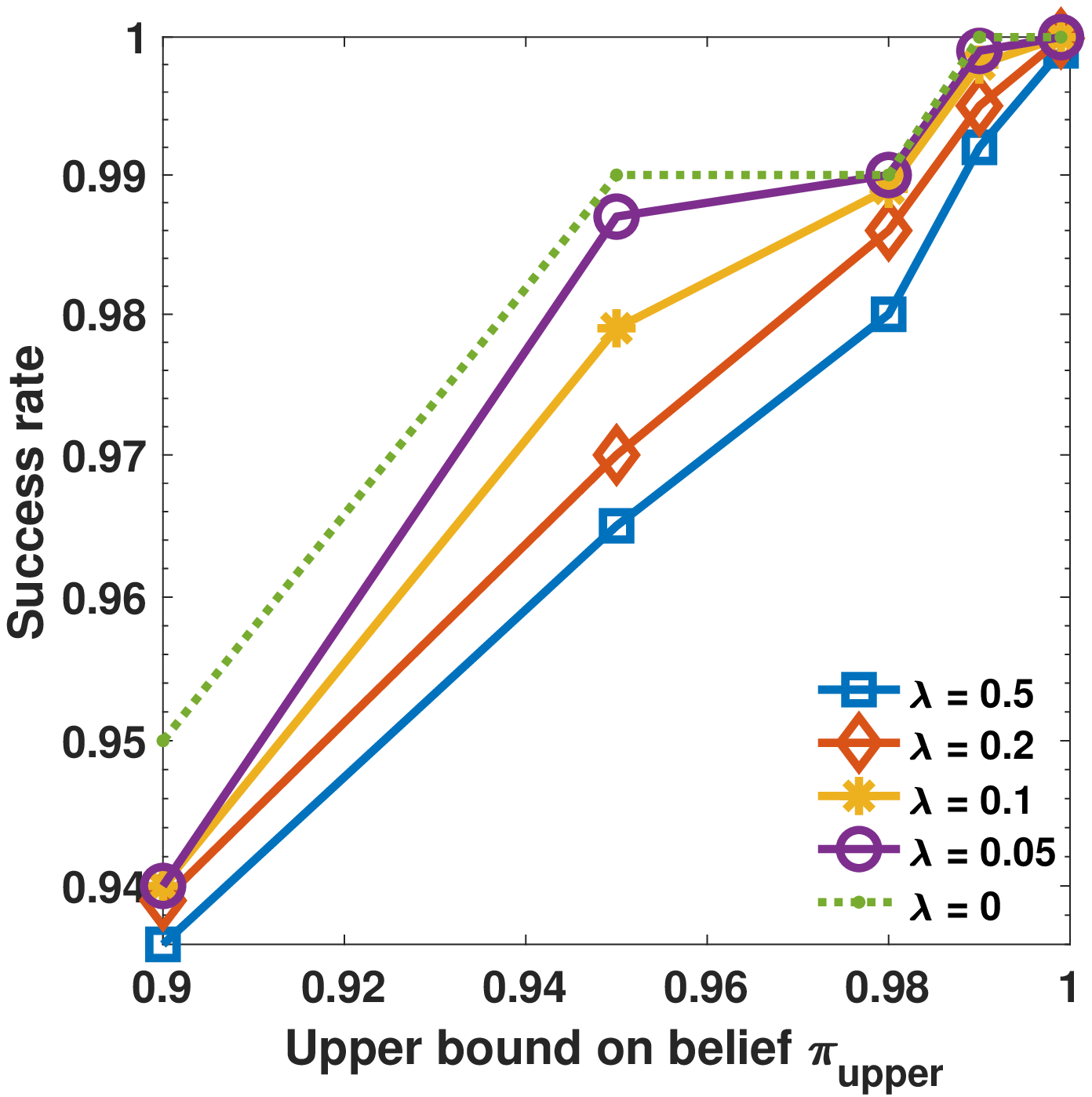}
\label{fig:success1}}
\caption{Success ratio performance of the algorithm when $\pi_{\mathrm{upper}}$ is varied from 0.9 to 0.999.}
\label{fig:success}
\end{center}
\end{figure*}

\begin{figure*}
\begin{center}
\hspace{-0.8cm}
\subfloat[$\rho=0$]
{\includegraphics[width=6.8cm]{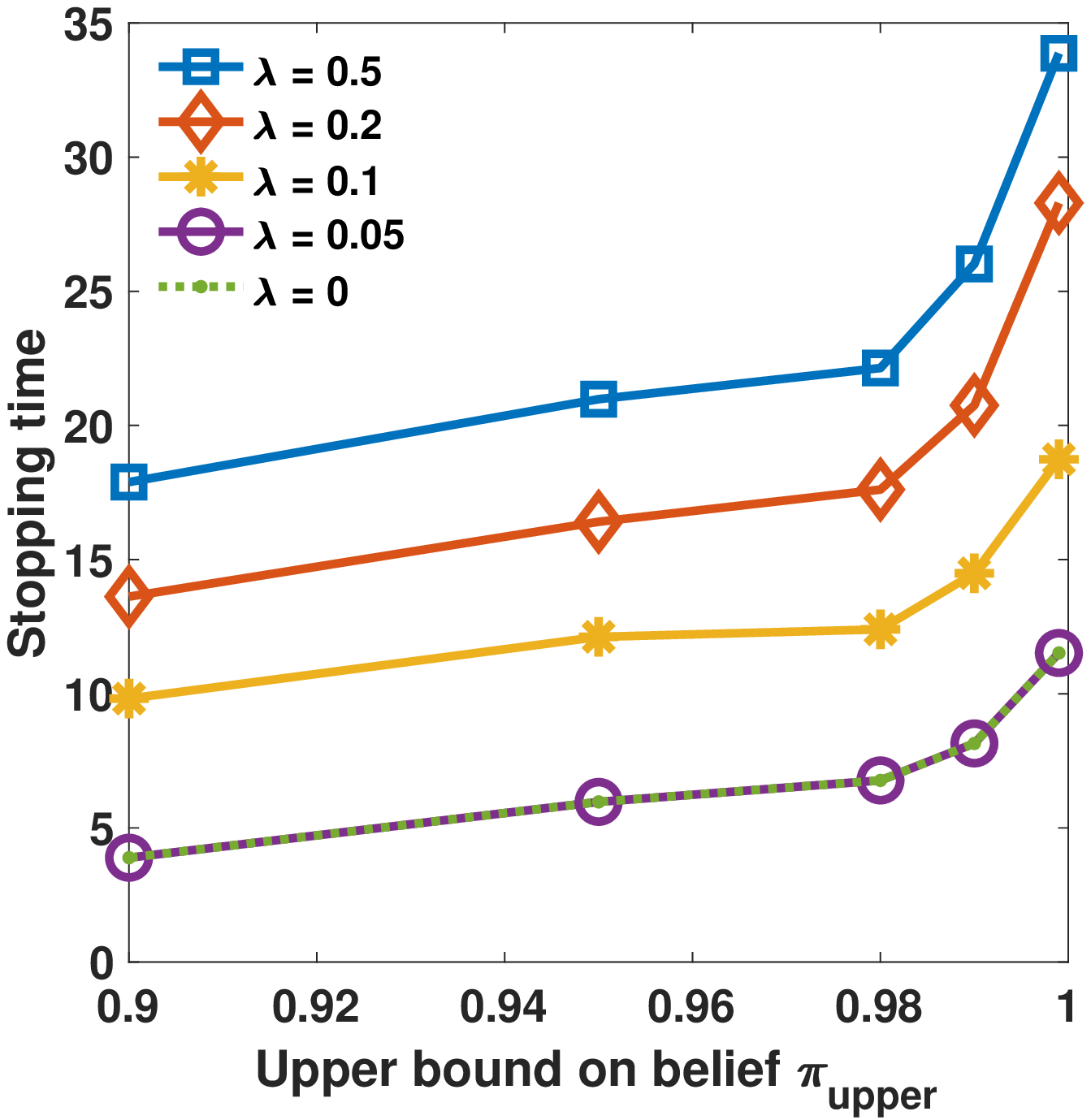}
\label{fig:delay0}}
\hspace{-1.4cm}
\subfloat[$\rho=0.3$]
{\includegraphics[width=6.8cm]{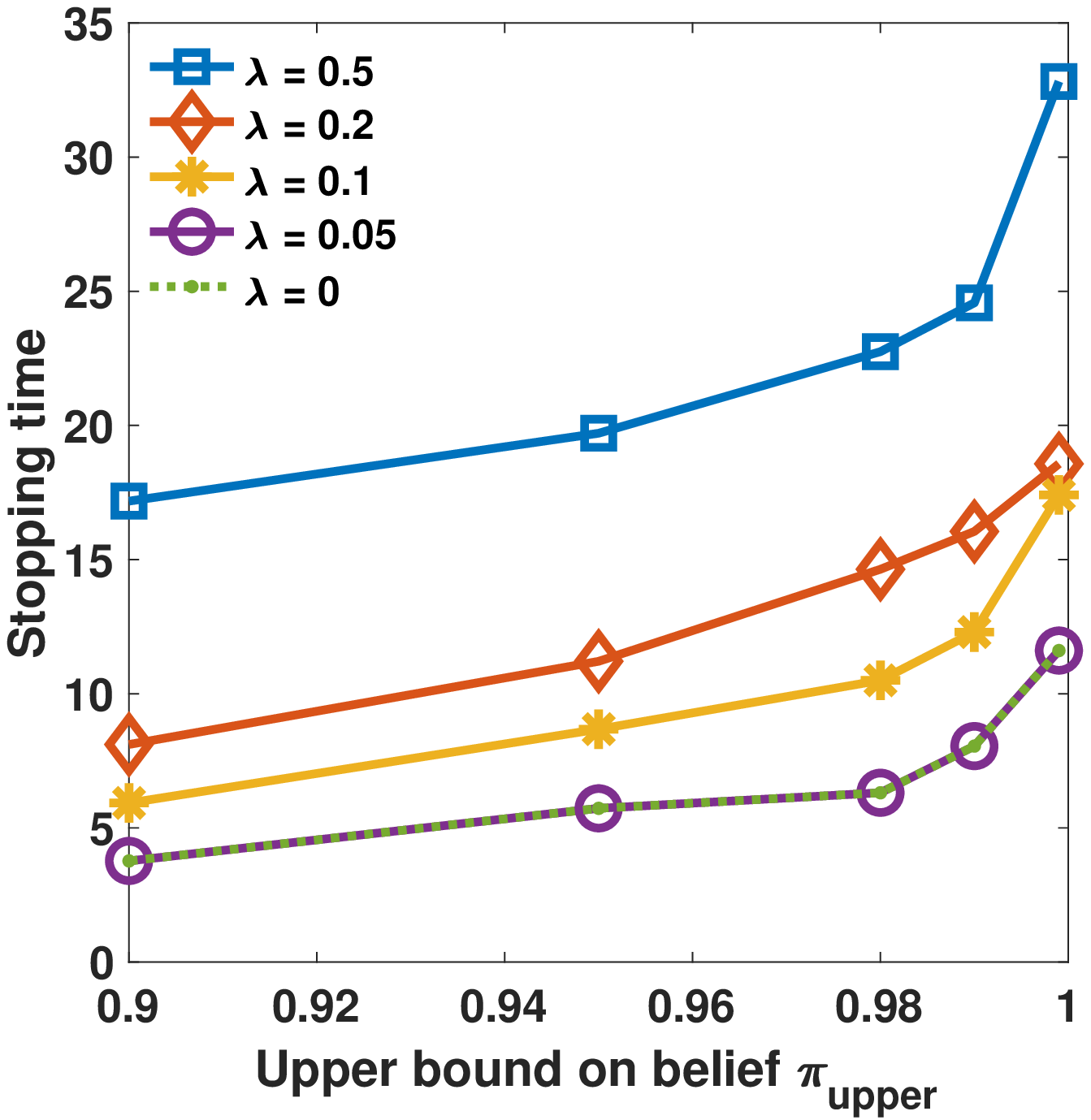}
\label{fig:delay3}}
\hspace{-1.4cm}
\subfloat[$\rho=1$]
{\includegraphics[width=6.8cm]{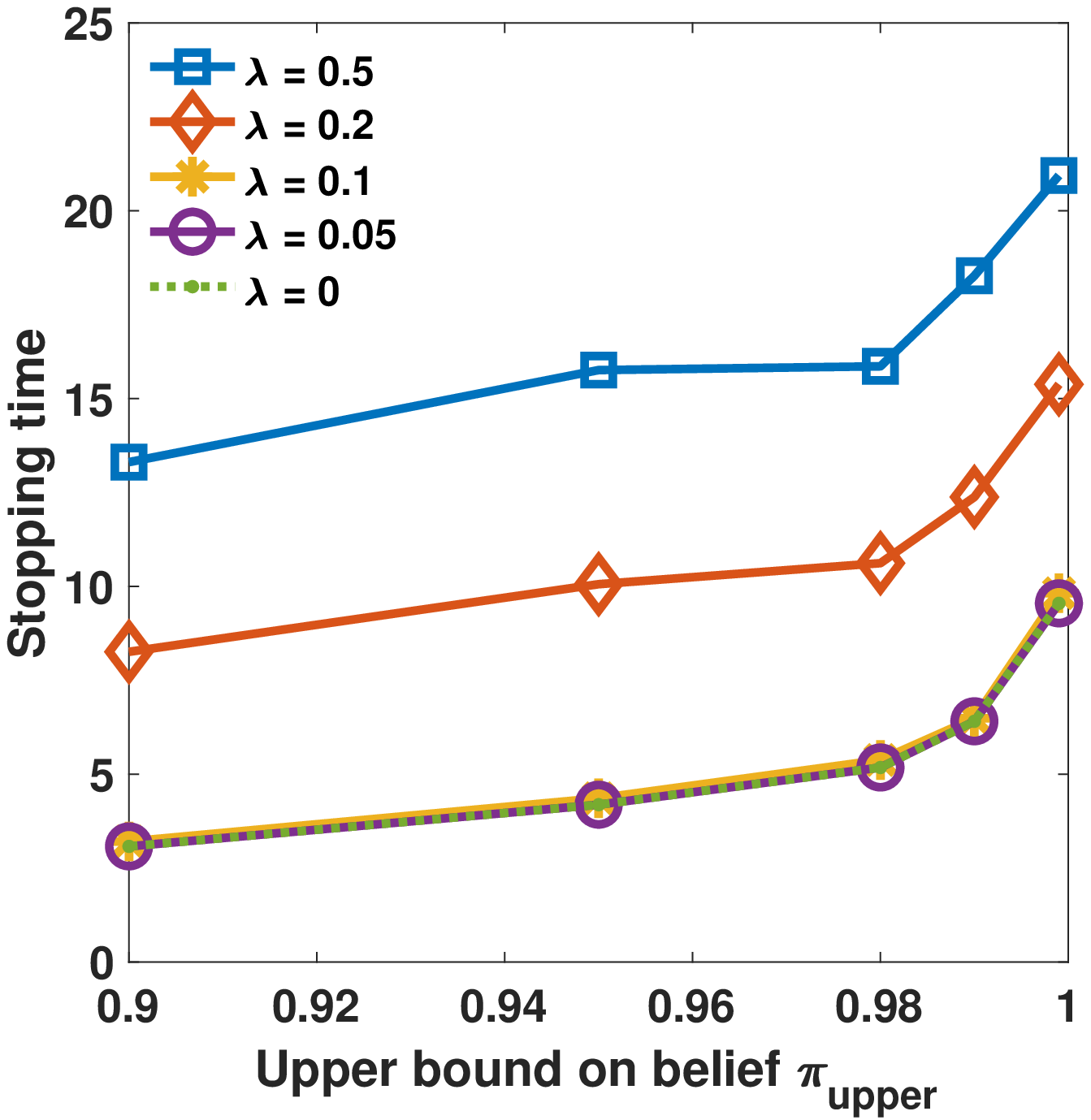}
\label{fig:delay1}}
\caption{Stopping time performance of the algorithm when $\pi_{\mathrm{upper}}$ is varied from 0.9 to 0.999.}
\label{fig:delay}
\end{center}
\end{figure*}


We set the number of sensors $N=3$, and therefore, the number of hypotheses is $8$, and the number of possible actions is $7$. The crossover probability of the sensor measurements is assumed to be $p=0.8$. The maximum number of time slots for every episode (trial or run) is taken as $T_{\max}=300$.
\begin{itemize}
\item \emph{Dependence model for the processes:} The two processes indexed by $1$ and $2$ are assumed to be dependent and the third process is independent of the other two. The probability of a process being normal is taken as $q=0.8$. Further, the correlation between the dependent processes is denoted by $\rho\in[0,1]$. Thus, the joint probability mass function of the two dependent processes is given below:
{\begin{align}
\bbP\lc \vecs_1=0,\vecs_2=0\rc &= q^2+\rho q(1-q)\\
\bbP\lc \vecs_1=0,\vecs_2=1\rc &= q(1-q)(1-\rho)\\
\bbP\lc \vecs_1=1,\vecs_2=0\rc &= q(1-q)(1-\rho)\\
\bbP\lc \vecs_1=1,\vecs_2=1\rc &= (1-q)^2+\rho q(1-q).
\end{align}}
\item \emph{Actor-critic framework:} Both neural networks (corresponding to the actor and the critic) are assumed to have 3 layers each, and the activation function is ReLU for all the layers except the output layer of the actor. For the output layer, we use the softmax function as the activation function so that the probabilities add up to one. For a given posterior probability $\pi$, the actor chooses the action that has the maximum probability of obtaining a good reward. The learning rates for the actor and the critic are chosen as $0.0005$ and $0.005$, respectively. The value of $\gamma$ is taken as $0.9$. During the training phase, we used $1500$ episodes,  and for each episode, we randomly generated the true hypothesis according to the above given prior (parameterized by $q=0.8$ and $\rho\in[0,1]$).  In each episode, the neural networks are trained based on data over 100 time slots.
\item \emph{Performance metrics:} We employ two performance metrics to evaluate the performance of the algorithms:
\begin{enumerate}
\item \emph{Success ratio:} The algorithm can fail in two ways: 
\begin{enumerate}
 \item the confidence level remains below the desired value for all the time slots from $1$ to $T_{\max}$
 \item the estimated hypothesis is not the correct one.
\end{enumerate} The ratio of the number of successful episodes to the total number of episodes is called the success ratio.
\item \emph{Stopping time:} The stopping time is defined as the average number of time slots required by the algorithm to decide on a hypothesis. While computing the average, we do not consider the case in which the confidence level is less than the desired value for the all time slots from $1$ to $T_{\max}$ and the algorithm does not make any decision. However, the cases where the algorithm decided on a wrong hypothesis are included in the computation of stopping time.
\end{enumerate}
\end{itemize}

We present our results in \Cref{fig:delay,fig:success} which plot the algorithm performance by varying the upper threshold on the belief $\pi_{\mathrm{upper}}$, the correlation coefficient $\rho$ and cost per sensor measurement $\lambda$.  Our observations from the results are as follows:
\begin{itemize}
\item \emph{Upper threshold on the belief $\pi_{\mathrm{upper}}$:} As $\pi_{\mathrm{upper}}$ increases, the hypothesis estimate becomes more accurate, and so the success ratio also increases, as shown  in \Cref{fig:success}. Further, to improve the accuracy, the algorithm requires more time slots to reach a conclusion on the states of the processes. Consequently, the stopping time also grows with $\pi_{\mathrm{upper}}$, as shown in \Cref{fig:delay}. This observation is in agreement with the intuition that as the stopping time increases, the algorithm receives more information about the processes, and as a result, the decisions become more accurate and are made faster.
\item \emph{Cost per sensor measurement $\lambda$:} From \Cref{fig:success}, we observe that the success ratios corresponding to different values of $\lambda$ are relatively close to each other, indicating that the sensor measurement cost does not have a significant impact on the success ratio. As $\lambda$ increases, the algorithm chooses fewer sensors per time instant and as a result, the stopping time grows, as shown in \Cref{fig:delay}. This behavior is expected from the nature of the instantaneous reward given in \eqref{eq:imm_reward}. This is  because  the algorithm tries to decrease the { last term } in \eqref{eq:imm_reward} by decreasing the number of sensors per time instant. The case of $\lambda=0$ implies no restriction on the number of sensors to be chosen at any given time instant. From our experiments, we find that when $\lambda=0$, the algorithm chooses all three sensors at all time instants. Since the algorithm in this case has the most information regarding the processes, decisions are made faster and more accurately. On the other hand, $\lambda=0.5$ corresponds to the most restrictive case in which the algorithm picked only one sensor per time instant.
\item \emph{Correlation coefficient $\rho$: } \Cref{fig:success} shows that the success ratio does not vary much when $\rho$ is varied while keeping $\lambda$ and $\pi_{\mathrm{upper}}$ constant.  On the contrary, from \Cref{fig:delay}, we see that the stopping time significantly depends on $\rho$. When $\rho=1$, the two dependent processes are identical. Thus, two measurements, one corresponding to one of the identical measurements, and the other corresponding to the independent process, contain the same information as that provided by three sensors. Hence, the plots corresponding to $\lambda=0.1,0.05$ and $0$ yield similar results. The slight improvement in the performance for $\lambda=0$ compared to that for the case when $\lambda=0.1$ is due to the impact of measurement noise. When $\lambda=0$, there are always two noisy measurements corresponding to the identical processes, and so the effective noise variance becomes smaller. The stopping time increases for higher values of $\lambda$, but we note that the stopping time corresponding to $\rho=1$ is the least among all the values of $\rho$ considered. Also, from \Cref{fig:delay}, we observe that as $\rho$ increases, the stopping time diminishes. This is because as $\rho$ increases, the mutual information between the sensor measurement corresponding to one of the dependent process and the state of the other dependent process increases. Thus, the algorithm requires a smaller number of sensors to estimate the hypothesis for the same level of confidence.  Hence, we conclude that our algorithm learns the underlying dependence pattern of the processes and changes its policy accordingly. This feature of the algorithm is an extra advantage of the algorithm apart from its low complexity, compared to other traditional model-based algorithms.
\end{itemize}
\section{Conclusion}
In this paper, we studied the anomaly detection problem where the goal is to identify an unknown number of anomalies among the given set of processes. We formulated the anomaly detection problem as a reward maximization problem based on a confidence metric using the expected Bayesian log-likelihood ratio and an additive penalty term that accounts for the cost of sensing. The reformulation led to an infinite-horizon, average-reward MDP over a finite-dimensional belief space. A low-complexity deep reinforcement learning algorithm based on the actor-critic framework was developed. Through numerical experiments, we studied the algorithm performance under various dependence patterns and different values of the sensing cost. The numerical results show that our algorithm is able to adapt according to any unknown statistical dependence between the processes. Deriving theoretical guarantees for the anomaly detection problem (MDP) using mathematical tools from information theory and statistical signal processing can be an interesting direction for future work.
\bibliographystyle{IEEEtran}
\bibliography{Supporting_Files/IEEEabrv,Supporting_Files/bibJournalList,Supporting_Files/AnomalyDetection}
\end{document}